\documentclass[a4paper]{article}

\usepackage{INTERSPEECH2020}
\usepackage{multirow, multicol}
\usepackage{arydshln}
\usepackage{url,times,booktabs}
\usepackage{mathtools}
\usepackage{cite}

\title{Improved RawNet with Feature Map Scaling for\\ Text-independent Speaker Verification using Raw Waveforms}
\name{Jee-weon Jung$^*$\thanks{$^*$Equal contribution}, Seung-bin Kim$^*$, Hye-jin Shim, Ju-ho Kim, Ha-Jin Yu$^\dag$\thanks{$^\dag$Corresponding author}}
\address{
  School of Computer Science, University of Seoul, Republic of Korea}
\email{jeewon.leo.jung@gmail.com, kimho1wq@naver.com, shimhz6.6@gmail.com, \\wngh1187@naver.com, hjyu@uos.ac.kr}

\begin{document}

\maketitle
\begin{abstract}
Recent advances in deep learning have facilitated the design of speaker verification systems that directly input raw waveforms. 
For example, RawNet \cite{jung2019rawnet} extracts speaker embeddings from raw waveforms, which simplifies the process pipeline and demonstrates competitive performance. 
In this study, we improve RawNet by scaling feature maps using various methods. 
The proposed mechanism utilizes a scale vector that adopts a sigmoid non-linear function. 
It refers to a vector with dimensionality equal to the number of filters in a given feature map. 
Using a scale vector, we propose to scale the feature map multiplicatively, additively, or both. 
In addition, we investigate replacing the first convolution layer with the sinc-convolution layer of SincNet. 
Experiments performed on the VoxCeleb1 evaluation dataset demonstrate the effectiveness of the proposed methods, and the best performing system reduces the equal error rate by half compared to the original RawNet.  
Expanded evaluation results obtained using the VoxCeleb1-E and VoxCeleb-H protocols marginally outperform existing state-of-the-art systems. 
\end{abstract}

\section{Introduction}
With the recent advances in deep learning, many speaker verification studies have replaced the acoustic feature extraction process with deep neural networks (DNNs) \cite{jung2019rawnet,muckenhirn2018towards,jung2018complete, ravanelli2018speaker}. 
In the preliminary stage of utilizing DNNs for speaker verification, acoustic features, e.g., Mel-frequency cepstral coefficients and Mel-filterbank energy features, were utilized as input to DNNs \cite{d-vector, snyder2018x, okabe2018attentive, jung2019spatial}. 
In contrast, many recent studies have also used less processed features, e.g., spectrograms and raw waveforms \cite{hajibabaei2018unified, nagrani2020voxceleb, jung2018avoiding22, ravanelli2019sincnet}, hypothesizing that the usage of such less processed features as input allows data-driven approaches with DNNs to yield better discriminative representations compared to using knowledge-based acoustic features. 
Following this trend, many recent systems, e.g., RawNet \cite{jung2019rawnet}, have demonstrated competitive results using raw waveforms as input for speaker verification. 

The attention mechanism was initially designed to emphasize more important elements in sequence-to-sequence processing \cite{bahdanau2014neural, vaswani2017attention, chan2016listen, zhu2018self, safari2019self}. 
It has been adopted to several tasks, including speaker verification. 
Among various methods, self-attentive pooling has been applied to speaker verification to aggregate frame-level representations into a single utterance-level representation \cite{zhu2018self}. 
Here, the term ``self’’ refers to a property of the attention mechanism that no external data, e.g., phoneme labels \cite{zhou2019cnn}, are used. 
Compared to conventional global average pooling, self-attentive pooling involves assigning a weight to each frame and conducting weighted summation. 
Recent attention mechanisms, e.g., multi-head self-attentive pooling, have also been investigated\cite{safari2019self}, and such methods have demonstrated further performance improvements. 
However, applying an attention mechanism to speaker verification is more focused on attentive pooling than applying such mechanisms to intermediate feature maps in image domain tasks \cite{okabe2018attentive, zhu2018self, safari2019self}.  

In this study, we propose to scale the filter axis of feature maps using a sigmoid-based mechanism, which we refer to as feature map scaling (FMS). 
The FMS uses a scale vector whose dimension is identical to the number of filters, where each value is between 0 and 1, similar to an attention map used for an attention mechanism, with the exception that a sigmoid activation function is employed rather than a softmax function. 
The underlying hypothesis of using sigmoid activation functions (rather than the softmax function) to independently perform scaling is that, differing from few other tasks, an attention mechanism that exclusively selects only a few filters may remove an excessive amount of discriminative information. 
In addition, in light of the recent successes of attentive pooling mechanisms in speaker verification tasks, we investigate replacing the gated recurrent unit (GRU) layer of RawNet, which performs aggregation of frame-level representations, with the self-attentive pooling and self-multi-head-attentive pooling mechanisms.  

Specifically, we propose to apply the FMS to feature maps by either multiplying, adding, or performing both sequentially, as shown in Figure \ref{fig:frm}. 
By multiplicatively scaling a feature map, we expect to emphasize each filter of a feature map independently. 
In addition, by applying an FMS through adding, we expect to provide small perturbations that lead to increased discriminative power. 
This is inspired by a previous study \cite{zhang2018vector} that showed analyzing small alterations in high-dimensional space can drastically change discriminative power. 
By hypothesizing that these two approaches function in a complementary manner, we also propose to apply both approaches in sequence. 
In experiments, the proposed methods were applied to the output feature maps of each residual block following the literature \cite{woo2018cbam, hu2018squeeze}. 
In addition, we investigated replacing RawNet’s first convolutional layer with a sinc-convolution layer \cite{ravanelli2018speaker}, which has been reported to better capture aggregated frequency responses than the conventional convolutional layer.  

The remainder of this paper is organized as follows. 
Section \ref{sec:rawnet} describes the RawNet system, which we use as a baseline with several modifications. In Section \ref{sec:proposed}, we introduce the proposed FMS. 
Section \ref{sec:exp} discusses experimentation and presents an analysis of the experimental results. 
Finally, conclusions are presented in Section \ref{sec:conclusion}. 
\section{Baseline composition: RawNet}
\label{sec:rawnet}
RawNet is a neural speaker embedding extractor that inputs raw waveforms directly without preprocessing techniques and outputs speaker embeddings designed for speaker verification \cite{jung2019rawnet}. 
The underlying assumption behind using a DNN is that speaker embeddings extracted directly from raw waveforms by replacing an acoustic feature extraction with more hidden layers are expected to yield more discriminative representations as the amount of available data increases. 
RawNet adopts a convolutional neural network-gated recurrent unit (CNN-GRU) architecture, in which the first CNN layer has stride size identical to the filter length. 
The front CNN layers comprise residual blocks followed by a max-pooling layer, and extracts frame-level representations. 
Then a GRU layer aggregates frame-level features into an utterance-level representation, which is the final timestep of the GRU's output. 
The GRU layer is then connected to a fully-connected layer, where its output is used as the speaker embedding. 
Finally, the output layer receives a speaker embedding and performs identification in the training phase. 

\begin{table}[t]
 \caption{DNN architecture of the proposed System (referred to as \textit{RawNet2} for brevity). An output layer conducts speaker identification in the training phase, and is removed after the training. BN and LeakyReLU at the beginning of the first block are omitted following \cite{he2016identity}. Numbers denoted in Conv and Sinc-conv refers to filter length, stride, and number of filters. This architecture can also be used to represent the baseline (``\#3-ours” in Table \ref{tab:baseline}) by removing FMS operations and using a convolutional layer instead of a sinc-conv layer.}
  \centering
  \label{tab:DNN_arch}
  \begin{tabular}{l c c}
  \toprule
  \textbf{Layer} & \textbf{Input:59049 samples} & \textbf{Output shape}\\
  \toprule
   & Sinc(251,1,128) & \multirow{3}{*}{(19683, 128)}\\
  Sinc & MaxPool(3)\\
  -conv& BN & \\
  & LeakyReLU & \\
  \midrule
  Res block & 
    $\left \{
      \begin{tabular}{c}
      BN \\
      LeakyReLU\\
      Conv(3,1,128)\\
      BN\\
      LeakyReLU\\
      Conv(3,1,128)\\
      \hdashline
      MaxPool(3)\\
      FMS\\
      \end{tabular}
    \right \}$
    $\times$2
    
  & (2187, 128)\\
  \midrule
  Res block & 
    
    $\left \{
      \begin{tabular}{c}
      BN \\
      LeakyReLU\\
      Conv(3,1,256)\\
      BN\\
      LeakyReLU\\      
      Conv(3,1,256)\\
      \hdashline
      MaxPool(3)\\
      FMS\\
      \end{tabular}
    \right \}$ 
    $\times$4
  & (27, 256)\\
  \midrule
  GRU & GRU(1024) & (1024,)\\
  \midrule
  Speaker & \multirow{2}{*}{FC(1024)} & \multirow{2}{*}{(1024,)}\\
  embedding & & \\
  \bottomrule
  \end{tabular}
\end{table}

To construct the baseline used in this study, we implemented several modifications to the original RawNet. First, we modified the structure of the residual blocks to a pre-activation structure \cite{he2016identity}. 
Second, we simplified the loss functions from using categorical cross-entropy (CCE), center \cite{wen2016discriminative}, and speaker basis loss \cite{heo2019end} to using only CCE loss. 
Third, we omitted a CNN pretraining scheme. 
Fourth, we modified the training dataset from VoxCeleb1 to VoxCeleb2 to utilize recently expanded evaluation protocols that consider the entire VoxCeleb1 dataset. 
Finally, we applied a test time augmentation (TTA) method in the evaluation phase \cite{Voxceleb2} to extract multiple speaker embeddings from a single utterance by cropping with overlaps where the duration is identical to that in the training phase. 
Then, the average of the speaker embeddings is used as the final speaker embedding. 
Through these modifications, we achieve a relative error reduction (RER) of 37.50 \%. 

\section{Filter-wise feature map scaling}
\label{sec:proposed}
\begin{figure}[t]
  \centering
  \includegraphics[width=\linewidth]{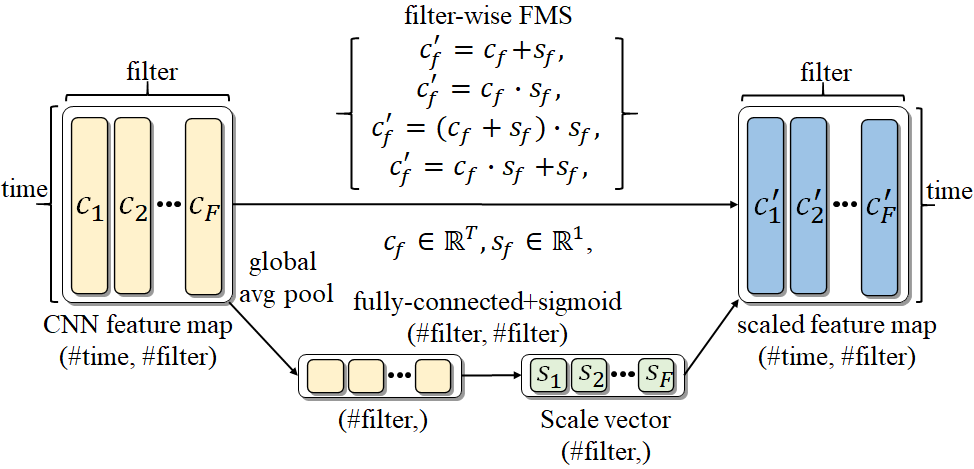}
  \caption{Illustration of the four methods using the proposed FMS. Here, $s_f$ is broadcasted to perform element-wise calculations with $c_f$.}
  \label{fig:frm}
\end{figure}

We propose to independently scale each filter of a feature map using a filter-wise feature map scaling (FMS) technique. 
The FMS uses a scale vector whose dimension is identical to the number of filters with values between 0 and 1 derived using sigmoid activation. 
Its purpose is to independently modify the scale of a given feature map, i.e., the output of a residual block, to derive more discriminative representations. 
We also propose various methods to utilize the FMS to scale given feature maps, i.e., multiplication, addition, and applying both. 
Note that these proposed approaches do not require additional hyperparameters. 

Here, let $c = [c_1, c_2, \dotsb, c_F]$ be a feature map of a residual block, i.e., $c_f \in \mathbb{R}^{T}$, where $T$ is the sequence length in time, and $F$ is the number of filters. 
We derive a scale vector to conduct FMS by first performing global average pooling on the time axis, and then feed-forwarding through a fully-connected layer followed by sigmoid activation. 
By expressing a scale vector as $s = [s_1, s_2, \dotsb, s_F]$, i.e., $s_f \in \mathbb{R}^{1}$, we first propose to derive a scaled feature map $c'=[c'_1, c'_2, \dotsb, c'_F]$, i.e., $c'_f \in \mathbb{R}^{T}$, to scale the feature map in an additive method expressed as follows: 
\begin{equation} 
    c'_f = c_f+s_f,
\end{equation} 
where $s_f$ is broadcasted, i.e. copied, to perform element-wise calculation. 
We also propose to scale the feature map in a multiplicative method: 
\begin{equation} 
    c'_f = c_f \cdot s_f. 
\end{equation} 
These two methods can be applied sequentially, where either method can be performed first, expressed as follows:
\begin{equation} 
    \label{eq:frm_add_mul} 
    c'_f = (c_f + s_f) \cdot s_f, 
\end{equation} 
\begin{equation} 
    \label{eq:frm_mul_add} 
    c'_f = c_f \cdot s_f+s_f. 
\end{equation} 
We also propose to use two individual scale vectors, i.e., one for addition, and the other for multiplication for \eqref{eq:frm_mul_add}, because it can be interpreted as $c'_f = (c_f +1) \times s_f$. 
Figure \ref{fig:frm} shows the proposed methods using FMS to scale feature maps. 
Here, we applied the propose methods using the FMS to the outputs of residual blocks in the baseline system following the literature \cite{woo2018cbam, hu2018squeeze}. 

The proposed method using multiplicative FMS for scaling has commonality with the widely used attention mechanism \cite{bahdanau2014neural, vaswani2017attention, chan2016listen} applied in the filter domain, which exclusively emphasizes a given feature map using a softmax activation. 
This can be interpreted as using the recently proposed multi-head attention mechanism \cite{vaswani2017attention} in the filter domain, where the number of the heads is equal to the number of filters. 
We apply scaling using a sigmoid function rather than exclusively performing scaling using a softmax function because information might be removed excessively when a conventional softmax-based attention mechanism. 
In translation or image classification tasks, performing exclusive concentration is reasonable; however, we hypothesize that different filters would yield complementary features for speaker verification, thereby making independent scaling more adequate.  

The proposed method with additive FMS for filter-wise scaling adds a value between 0 and 1 to a given feature map. 
The purpose is to apply data-driven perturbation to a feature map with a relatively small value. 
Here, it is assumed that this may increase the discriminative power of the feature maps. 
This concept is inspired by a phenomenon demonstrated by Zhang \textit{et. al.} \cite{zhang2018vector}, where the discriminative power of a DNN's high-dimensional intermediate representation can differ significantly with small perturbations. 
In addition, we assume that applying an additive FMS combined with a multiplicative FMS will lead to further improvements.  

We also investigated replacing the RawNet’s first convolutional layer with a sinc-convolution (sinc-conv) layer, which was first proposed to process raw waveforms by performing time-domain convolutions \cite{ravanelli2018speaker, ravanelli2019sincnet}. 
It is a type of a bandpass filter, where cut-off frequencies are set as parameters that are optimized with other DNN parameters. 
With fewer parameters, i.e., $2 \times \#filter$, the sinc-conv layer is frequently employed in DNNs that directly input a raw waveform. 
Table \ref{tab:DNN_arch} details the overall architecture of the proposed system.  

\begin{table}[t]
  \caption{Performance comparison according to modifications of the baseline construction ($^*$: data augmentation). Equal error rate (EER) is reported using the original VoxCeleb1 evaluation dataset. \textit{Ours} shows the results of applying identity mapping \cite{he2016identity}, modifying the dimensionality of the code representation, and increasing the training set.}
  \label{tab:baseline}
  \centering
  \begin{tabular}{lcccc}
    \toprule
    System & Trained on & TTA &EER & RER\\
    \midrule
    i-vector \cite{shon2018frame} & Vox1 & $\times$ & 5.40 & -\\
    specCNN* \cite{hajibabaei2018unified}& Vox1 & $\times$ & 4.30 & -\\
    x-vector* \cite{snyder2018x} & Vox2 & $\times$ & 3.10 & -\\
    \midrule
    \#1-RawNet \cite{jung2019rawnet} & Vox1 & $\times$ &4.80 & -\\
    \#2-Ours & Vox2 & $\times$ & 3.52 & 26.67\\
    \#3-\textbf{Ours} & Vox2 & $\circ$ & \textbf{3.00} & 37.50\\
    \bottomrule
  \end{tabular}
\end{table}
\begin{table}[t]
  \caption{Various applications of the proposed FMS. \textit{Baseline} refers to the modified RawNet (Table \ref{tab:baseline}). Mechanism addresses variations of applying the proposed method. $c$ is the feature map, and $s$ is the scale vector used to conduct FMS derived from $c$. ``sep’’ indicates using \textbf{sep}arate scale vectors for additive and multiplicative scaling. Performance is reported in terms of EER and RER.}
  \label{tab:filt_wise_scale}
  \centering
  \begin{tabular}{lcccc}
    \toprule
    System & Mechanism  &EER & RER\\
    \midrule
    Baseline & - & 3.00 & -\\
    \midrule
    \#4-att & - & 2.89 & 3.67\\
    \#5-multi-att & - & 3.42 & -\\
    \#6-SE & - & 2.65 & 11.67\\
    \#7-CBAM & - & 2.89 & 3.67\\
    \midrule
    \#8-add & $c+r$ & 2.82 & 6.00\\
    \#9-mul & $c\times r$ & 2.66 & 11.33\\
    \#10-add-mul & $(c+r) \times r$ & 2.60 & 13.33\\
    \#\textbf{11-mul-add} & $c\times r + r$ & \textbf{2.56} & \textbf{14.67}\\
    \midrule
    \#12-mul-add-sep & $c\times r_1 + r_2$ &  2.57 & 14.33\\
    \bottomrule
  \end{tabular}
\end{table}
\begin{table}[t]
  \caption{Experimental results of replacing the first strided convolution layer with varying sinc-conv layer length proposed in SincNet \cite{ravanelli2018speaker, ravanelli2019sincnet}. Applied to System \#11-mul-add of Table \ref{tab:filt_wise_scale}.}
  \label{tab:sincconv}
  \centering
  \begin{tabular}{lcccc}
    \toprule
    System & Sinc-conv length  & EER & RER\\
    \midrule
    \#11-mul-add & - & 2.56 & -\\
    \midrule
    \#13 & 125 & 2.53 & 1.17\\
    \#14 & 195 & 2.54 & 0.78\\
    \textbf{\# 15-RawNet2} & \textbf{251} & \textbf{2.48} & \textbf{3.12}\\
    \#16 & 313 & 2.70 & -\\
    \#17 & 375 &  2.75 & -\\
    \bottomrule
  \end{tabular}
\end{table}

\begin{table*}[t]
 \caption{Results of comparison to state-of-the-art systems on expanded VoxCeleb1-E and VoxCeleb-H evaluation protocols.}
  \centering
  \label{tab:sota}
  \begin{tabular}{l c c c c c c c }
  \toprule
    & Input Feature & Front-end & Aggregation & Loss & Dims & EER (\%) \\
  \toprule 
  \textbf{VoxCeleb1-E}\\
  \midrule\midrule
  Chung \textit{et. al.} \cite{Voxceleb2} & Spectrogram & ResNet-50 & TAP & Softmax+Contrastive & 512 & 4.42\\
  Xie \textit{et. al.} \cite{xie2019aggregation} & Spectrogram & Thin ResNet-34 & GhostVLAD & Softmax & 512 & 3.13\\
   Nagrani \textit{et al.} \cite{nagrani2020voxceleb} & Spectrogram & Thin-ResNet-34 & GhostVLAD & Softmax & 512 & 2.95\\
  \textbf{Ours} & Raw waveform & RawNet2 & GRU & Softmax & 1024  & \textbf{2.57}\\

  \bottomrule
  \textbf{VoxCeleb1-H}\\
  \midrule\midrule
  Chung \textit{et. al.} \cite{Voxceleb2} & Spectrogram & ResNet-50 & TAP & Softmax+Contrastive & 512 & 7.33\\
  Xie \textit{et. al.} \cite{xie2019aggregation} & Spectrogram & Thin ResNet-34 & GhostVLAD & Softmax & 512 &  5.06\\
   Nagrani \textit{et al.} \cite{nagrani2020voxceleb} & Spectrogram & Thin-ResNet-34 & GhostVLAD & Softmax & 512 &  4.93\\
  \textbf{Ours} & Raw waveform & RawNet2 & GRU & Softmax & 1024 & \textbf{4.89}\\
  \bottomrule
  \end{tabular}
\end{table*}

\section{Experiments and result analysis}
\label{sec:exp}
All experiments reported in this paper were conducted using PyTorch \cite{paszke2019PyTorch}, and the code is available at \url{https://github.com/Jungjee/RawNet}.

\subsection{Dataset}
We used the VoxCeleb2 dataset \cite{Voxceleb2} for training, and we used the VoxCeleb1 dataset \cite{Voxceleb} to evaluate various protocols. 
The VoxCeleb2 dataset contains over one million utterances from 6112 speakers, and the VoxCeleb1 dataset contains approximately 330 hours of recordings from 1251 speakers for text-independent scenarios. Both datasets were obtained automatically from YouTube. Note that the VoxCeleb2 dataset is an extended version of the VoxCeleb1 dataset. 

\subsection{Experimental configurations}
We used raw waveforms with pre-emphasis applied as input to the DNNs \cite{jung2018complete, jung2018avoiding22, jung2019rawnet}. 
For the experiments in which the first convolutional layer was replaced with sinc-conv layer, we followed the literature \cite{ravanelli2018speaker}. 
This study did not apply pre-emphasis but performed layer normalization \cite{ba2016layer} to raw waveform. 
Here, we modified the duration of the input waveforms to $59049$ samples ($\approx$ 3.69 s) in the training phase to facilitate mini-batch construction. 
In the testing phasing, we applied TTA with a $20$ \% overlap. 

We used Leaky ReLU activation functions \cite{leaky} with a negative slope of $0.3$ following implementations of \cite{keras}. 
The dimension of the speaker embedding is $1024$. 
The AMSGrad optimizer \cite{reddi2019convergence} with a learning rate of $0.001$ was used. 
A weight decay with $\lambda=1e^{-4}$ was applied. We used CCE loss as the objective function. 
The other parameters related to the system architecture are described in Table \ref{tab:DNN_arch} and the literature \cite{jung2019rawnet}.

\subsection{Results analysis}
Table \ref{tab:baseline} shows the performance according to the modifications made to the RawNet system (Section \ref{sec:rawnet}) using the original VoxCeleb1 evaluation set. 
Here, the top three rows describe existing systems using the same dataset for comparison. 
The results indicate that the original RawNet demonstrates competitive performance. 
For x-vectors, we show the results for an improved version reported in the literature \cite{nagrani2020voxceleb}. 
Here, System \#1 describes the performance of RawNet \cite{jung2019rawnet}, and System \#2 shows the result of changing the DNN architecture and expanding the training set to the VoxCeleb2 dataset from System \#1. 
System \#3 shows the results obtained by applying TTA to System \#2. 
As can be seen, the results demonstrate that the applied changes were effective and resulted in RER of 37.5 \% compared to the original RawNet. 
Note that we used System \#3 as the baseline in all experiments.  

Systems \#4 through \#7 of Table \ref{tab:filt_wise_scale} show the results obtained by applying various related methods. 
Systems \#4 and \#5 show the results obtained using the attention and multi-head attention mechanisms using the softmax-based exclusive attention map on the filter domain. 
Attention demonstrated marginal improvement, and multi-head attention reduced the performance matching the hypothesis discussed in Section \ref{sec:proposed}. 
Systems \#6 and \#7 describe the results of applying squeeze-excitation \cite{hu2018squeeze} and convolutional block attention module \cite{woo2018cbam} to the baseline. 
In addition, the experimental results obtained by replacing the GRU layer with self-attentive pooling or self-multi-head-attentive pooling reduced performance. 
These results demonstrate that, in the case of RawNet, the GRU better aggregates frame-level representations into an utterance-level representation. 
Among application of various related methods, System \#6 demonstrated the best result. 

Systems \#8 through \#12 of Table \ref{tab:filt_wise_scale} show the results of proposed FMS method with different configurations. 
Here, the ``Mechanism’’ column shows how we performed the proposed FMS method. 
System \#8 and System \#9 applied the two proposed methods, and \#10 and \#11 applied both methods at the same time in different order. 
The results show that the proposed methods yielded improvements with RERs of 6.00 \% and 11.33 \%. 
Applying both methods simultaneously further improved the performance, and System \#11 demonstrated an EER of 2.56 \%. 
System \#12 shows the result obtained using separate scale vectors for additive and multiplicative FMS. 
As shown, additional improvements were not observed.  

Table \ref{tab:sincconv} shows the results obtained by replacing the RawNet’s first convolutional layer with the sinc-conv layer of SincNet. 
Here, we used System \#11 to perform these comparative experiments. 
The result demonstrates that it provides 3.12 \% additional improvement in System \#15. 
However, the performance was easily affected by the length of the sinc-conv layer, i.e., setting an overly long filter length reduced performance. 
In the following, we refer to System \#15 that demonstrates the best performance as `RawNet2' for brevity. 
RawNet2 demonstrates an RER of 48.33 \% compared to the original RawNet (System \#1), thereby nearly halving the EER.   

Finally, Table \ref{tab:sota} compares the results obtained in various recent studies using the expanded evaluation protocols, i.e., VoxCeleb1-E and VoxCeleb1-H, which utilize more than 1000 speakers and 500000 trials compared to 40 speakers and 38000 trials in the original evaluation protocol.\footnote{We report all performance values using the \textit{cleaned} protocol.} 
The results show that the proposed RawNet2 marginally outperformed the state-of-the-art performance, i.e., EER of 2.87 \% for the VoxCeleb-E protocol and 4.89 \% for the VoxCeleb-H protocol. 
From the various experimental results given through Tables \ref{tab:baseline} to \ref{tab:sota}, we conclude that the proposed RawNet2 using the FMS demonstrates competitive performance despite its simple process pipeline of inputting raw waveforms to a DNN and measuring cosine similarity using the output speaker embeddings. 

\section{Conclusion}
\label{sec:conclusion}
In this paper, we have proposed various FMS-based methods to improve the existing RawNet system, which is neural speaker embedding extractor in which speaker embeddings are extracted directly from raw waveforms. The FMS uses a scale vector to perform scaling, where the dimension of the scale vector is identical to the number of filters. 
The FMS-based methods scale filters in feature maps to construct improved feature maps that focus on more important features in the frame-level feature map through addition, multiplication, or both. 
We applied various FMS-based methods to the output of each residual block. 
In addition, by replacing the first convolution layer with a sinc-conv layer, we achieved further improvements. 
The results of the evaluation performed using the original VoxCeleb1 protocol demonstrate an EER of 2.46 \%, while the original RawNet reported EER of 4.80 \%. 
In an evaluation using recently expanded evaluation protocols, the proposed method marginally outperformed the current state-of-the-art methods.  


\bibliographystyle{IEEEtran}
\bibliography{refs}

\begin{thebibliography}{10}
\providecommand{\url}[1]{#1}
\csname url@samestyle\endcsname
\providecommand{\newblock}{\relax}
\providecommand{\bibinfo}[2]{#2}
\providecommand{\BIBentrySTDinterwordspacing}{\spaceskip=0pt\relax}
\providecommand{\BIBentryALTinterwordstretchfactor}{4}
\providecommand{\BIBentryALTinterwordspacing}{\spaceskip=\fontdimen2\font plus
\BIBentryALTinterwordstretchfactor\fontdimen3\font minus
  \fontdimen4\font\relax}
\providecommand{\BIBforeignlanguage}[2]{{%
\expandafter\ifx\csname l@#1\endcsname\relax
\typeout{** WARNING: IEEEtran.bst: No hyphenation pattern has been}%
\typeout{** loaded for the language `#1'. Using the pattern for}%
\typeout{** the default language instead.}%
\else
\language=\csname l@#1\endcsname
\fi
#2}}
\providecommand{\BIBdecl}{\relax}
\BIBdecl

\bibitem{jung2019rawnet}
J.-w. Jung, H.-S. Heo, J.-h. Kim, H.-j. Shim, and H.-J. Yu, ``Rawnet: Advanced
  end-to-end deep neural network using raw waveforms for text-independent
  speaker verification,'' \emph{Proc. Interspeech 2019}, pp. 1268--1272, 2019.

\bibitem{muckenhirn2018towards}
H.~Muckenhirn, M.~Doss, and S.~Marcell, ``Towards directly modeling raw speech
  signal for speaker verification using cnns,'' in \emph{2018 IEEE
  International Conference on Acoustics, Speech and Signal Processing
  (ICASSP)}.\hskip 1em plus 0.5em minus 0.4em\relax IEEE, 2018, pp. 4884--4888.

\bibitem{jung2018complete}
J.~Jung, H.~Heo, I.~Yang, H.~Shim, and H.~Yu, ``A complete end-to-end speaker
  verification system using deep neural networks: From raw signals to
  verification result,'' in \emph{2018 IEEE International Conference on
  Acoustics, Speech and Signal Processing (ICASSP)}.\hskip 1em plus 0.5em minus
  0.4em\relax IEEE, 2018, pp. 5349--5353.

\bibitem{ravanelli2018speaker}
M.~Ravanelli and Y.~Bengio, ``Speaker recognition from raw waveform with
  sincnet,'' \emph{arXiv preprint arXiv:1808.00158}, 2018.

\bibitem{d-vector}
E.~Variani, X.~Lei, E.~McDermott, I.~L. Moreno, and J.~Gonzalez-Dominguez,
  ``Deep neural networks for small footprint text-dependent speaker
  verification,'' in \emph{International Conference on Acoustics, Speech and
  Signal Processing (ICASSP)}.\hskip 1em plus 0.5em minus 0.4em\relax IEEE,
  2014, pp. 4052--4056.

\bibitem{snyder2018x}
D.~Snyder, D.~Garcia-Romero, G.~Sell, D.~Povey, and S.~Khudanpur, ``X-vectors:
  Robust dnn embeddings for speaker recognition,'' in \emph{2018 IEEE
  International Conference on Acoustics, Speech and Signal Processing
  (ICASSP)}.\hskip 1em plus 0.5em minus 0.4em\relax IEEE, 2018, pp. 5329--5333.

\bibitem{okabe2018attentive}
K.~Okabe, T.~Koshinaka, and K.~Shinoda, ``Attentive statistics pooling for deep
  speaker embedding,'' \emph{arXiv preprint arXiv:1803.10963}, 2018.

\bibitem{jung2019spatial}
Y.~Jung, Y.~Kim, H.~Lim, Y.~Choi, and H.~Kim, ``Spatial pyramid encoding with
  convex length normalization for text-independent speaker verification,''
  \emph{Proc. Interspeech 2019}, pp. 4030--4034, 2019.

\bibitem{hajibabaei2018unified}
M.~Hajibabaei and D.~Dai, ``Unified hypersphere embedding for speaker
  recognition,'' \emph{arXiv preprint arXiv:1807.08312}, 2018.

\bibitem{nagrani2020voxceleb}
A.~Nagrani, J.~S. Chung, W.~Xie, and A.~Zisserman, ``Voxceleb: Large-scale
  speaker verification in the wild,'' \emph{Computer Speech \& Language},
  vol.~60, 2020.

\bibitem{jung2018avoiding22}
J.~Jung, H.~Heo, I.~Yang, H.~Shim, and H.~Yu, ``Avoiding speaker overfitting in
  end-to-end dnns using raw waveform for text-independent speaker
  verification,'' in \emph{Proc. Interspeech 2018}, 2018, pp. 3583--3587.

\bibitem{ravanelli2019sincnet}
M.~Ravanelli and Y.~Bengio, ``Learning speaker representations with mutual
  information,'' in \emph{Interspeech}, 2019.

\bibitem{bahdanau2014neural}
D.~Bahdanau, K.~Cho, and Y.~Bengio, ``Neural machine translation by jointly
  learning to align and translate,'' \emph{arXiv preprint arXiv:1409.0473},
  2014.

\bibitem{vaswani2017attention}
A.~Vaswani, N.~Shazeer, N.~Parmar, J.~Uszkoreit, L.~Jones, A.~N. Gomez,
  {\L}.~Kaiser, and I.~Polosukhin, ``Attention is all you need,'' in
  \emph{Advances in neural information processing systems}, 2017, pp.
  5998--6008.

\bibitem{chan2016listen}
W.~Chan, N.~Jaitly, Q.~Le, and O.~Vinyals, ``Listen, attend and spell: A neural
  network for large vocabulary conversational speech recognition,'' in
  \emph{2016 IEEE International Conference on Acoustics, Speech and Signal
  Processing (ICASSP)}.\hskip 1em plus 0.5em minus 0.4em\relax IEEE, 2016, pp.
  4960--4964.

\bibitem{zhu2018self}
Y.~Zhu, T.~Ko, D.~Snyder, B.~Mak, and D.~Povey, ``Self-attentive speaker
  embeddings for text-independent speaker verification.'' in
  \emph{Interspeech}, 2018, pp. 3573--3577.

\bibitem{safari2019self}
P.~Safari and J.~Hernando, ``Self multi-head attention for speaker
  recognition,'' \emph{Proc. Interspeech 2019}, pp. 4305--4309, 2019.

\bibitem{zhou2019cnn}
T.~Zhou, Y.~Zhao, J.~Li, Y.~Gong, and J.~Wu, ``Cnn with phonetic attention for
  text-independent speaker verification,'' in \emph{2019 IEEE Automatic Speech
  Recognition and Understanding Workshop (ASRU)}.\hskip 1em plus 0.5em minus
  0.4em\relax IEEE, 2019, pp. 718--725.

\bibitem{zhang2018vector}
J.~Zhang, N.~Inoue, and K.~Shinoda, ``I-vector transformation using conditional
  generative adversarial networks for short utterance speaker verification,''
  \emph{Proceedings of INTERSPEECH, Hyderabad, India}, 2018.

\bibitem{woo2018cbam}
S.~Woo, J.~Park, J.-Y. Lee, and I.~So~Kweon, ``Cbam: Convolutional block
  attention module,'' in \emph{Proceedings of the European Conference on
  Computer Vision (ECCV)}, 2018, pp. 3--19.

\bibitem{hu2018squeeze}
J.~Hu, L.~Shen, and G.~Sun, ``Squeeze-and-excitation networks,'' in
  \emph{Proceedings of the IEEE conference on computer vision and pattern
  recognition}, 2018, pp. 7132--7141.

\bibitem{he2016identity}
K.~He, X.~Zhang, S.~Ren, and J.~Sun, ``Identity mappings in deep residual
  networks,'' in \emph{European conference on computer vision}.\hskip 1em plus
  0.5em minus 0.4em\relax Springer, 2016, pp. 630--645.

\bibitem{wen2016discriminative}
Y.~Wen, K.~Zhang, Z.~Li, and Y.~Qiao, ``A discriminative feature learning
  approach for deep face recognition,'' in \emph{European conference on
  computer vision}.\hskip 1em plus 0.5em minus 0.4em\relax Springer, 2016, pp.
  499--515.

\bibitem{heo2019end}
H.-S. Heo, J.-w. Jung, I.-H. Yang, S.-H. Yoon, H.-j. Shim, and H.-J. Yu,
  ``End-to-end losses based on speaker basis vectors and all-speaker hard
  negative mining for speaker verification,'' \emph{arXiv preprint
  arXiv:1902.02455}, 2019.

\bibitem{Voxceleb2}
J.~S. Chung, A.~Nagrani, and A.~Zisserman, ``Voxceleb2: Deep speaker
  recognition,'' in \emph{Interspeech}, 2018.

\bibitem{shon2018frame}
S.~Shon, H.~Tang, and J.~Glass, ``Frame-level speaker embeddings for
  text-independent speaker recognition and analysis of end-to-end model,''
  \emph{arXiv preprint arXiv:1809.04437}, 2018.

\bibitem{xie2019aggregation}
W.~Xie, A.~Nagrani, J.~S. Chung, and A.~Zisserman, ``Utterance-level
  aggregation for speaker recognition in the wild,'' in \emph{ICASSP}, 2019.

\bibitem{paszke2019PyTorch}
A.~Paszke, S.~Gross, F.~Massa, A.~Lerer, J.~Bradbury, G.~Chanan, T.~Killeen,
  Z.~Lin, N.~Gimelshein, L.~Antiga \emph{et~al.}, ``Pytorch: An imperative
  style, high-performance deep learning library,'' in \emph{Advances in Neural
  Information Processing Systems}, 2019, pp. 8024--8035.

\bibitem{Voxceleb}
A.~Nagrani, J.~S. Chung, and A.~Zisserman, ``Voxceleb: a large-scale speaker
  identification dataset,'' in \emph{Interspeech}, 2017.

\bibitem{ba2016layer}
J.~L. Ba, J.~R. Kiros, and G.~E. Hinton, ``Layer normalization,'' \emph{arXiv
  preprint arXiv:1607.06450}, 2016.

\bibitem{leaky}
A.~L. Maas, A.~Y. Hannun, and A.~Y. Ng, ``Rectifier nonlinearities improve
  neural network acoustic models,'' in \emph{Proc. icml}, vol.~30, no.~1, 2013,
  p.~3.

\bibitem{keras}
F.~Chollet \emph{et~al.}, ``Keras,'' \url{https://github.com/keras-team/keras},
  2015.

\bibitem{reddi2019convergence}
S.~J. Reddi, S.~Kale, and S.~Kumar, ``On the convergence of adam and beyond,''
  \emph{arXiv preprint arXiv:1904.09237}, 2019.

\end{thebibliography}
\end{document}